\documentstyle[prl,aps]{revtex}
\begin{document}
\draft
\preprint{}
\title{Is there Ornstein-Zernike equation in the canonical ensemble?}
\author{J.A.\ White and S. Velasco}
\address{ Departamento de F\'{\i}sica Aplicada,  Facultad de Ciencias \\
Universidad de Salamanca, 37008 Salamanca, Spain}

\date{\today}
\maketitle
\begin{abstract} A general density-functional formalism using an extended
variable-space is presented  for classical fluids in the canonical ensemble
(CE). An exact equation is derived that plays the role of the Ornstein-Zernike
(OZ) equation in the grand canonical ensemble (GCE). When applied to the ideal
gas we obtain the exact result for the total correlation function $h_N$. For a
homogeneous fluid with $N$ particles the new equation only differs from OZ
by $1/N$ and it allows to obtain an approximate expression for $h_N$ in terms
of its GCE counterpart that agrees with the expansion of $h_N$ in powers of
$1/N$.
\end{abstract}


The  Ornstein-Zernike (OZ) equation is an essential ingredient of any theory of
the equilibrium microscopic structure of classical fluids since it provides an
exact relation between correlation functions at the two-particle level.  In the
world of the integral equation theory \cite{Hansen,Henderson92},  most
integral equations are based on this relation closed with an additional
approximate relation between the correlation functions. Closures like the
Percus-Yevick approximation, the hypernetted chain approximation, or the
mean spherical approximation, together with the OZ relation have led to
well-known  theories of classical fluids. In the world of density functional
theory (DFT)
\cite{Evans79,Evans92,Ashcroft95},  one makes explicit approximations for
the free-energy functional and, after using a variational principle, one
obtains  the direct correlation function via functional differentiation.
Inserting the direct correlation function into the OZ equation allows to obtain
the pair correlation function.  Furthermore,  as we shall see below, the OZ
equation is an identity that arises naturally from the DFT formalism
\cite{Evans79,Evans92,Ashcroft95}.

When using the OZ equation it is important to notice that it has been
formulated for the grand canonical ensemble (GCE) and thus care must be
taken when considering it in other statistical mechanics ensembles. Of
particular interest is the canonical ensemble (CE), not only from a
fundamental point of view but also  because of its relevance in the analysis of
systems with a fixed, finite number of particles $N$. The aim of this Letter is
to study the possibility of establishing an OZ relation in the CE
\cite{Hernando}. We shall see how the fixed-$N$ constraint precludes the
direct application of the GCE-OZ relation, this problem will be solved in a DFT
context by using an extended variable-space in which,  together with the
familiar inhomogeneous density,  a Lagrange multiplier is  considered as a
variable.

The OZ equation for a nonuniform system with fixed chemical potential
$\mu$ and temperature $T$, i.e., in the GCE, in the presence of an external
potential
$V_{\text{ext}}({\bf r})$, is usually expressed as the following relation
between the  total correlation function $h$ and the direct correlation function
$c^{(2)}$
\begin{equation} h({\bf r}_1,{\bf r}_2) = c^{(2)}({\bf r}_1,{\bf r}_2) +
\int \rho({\bf r}_3) c^{(2)}({\bf r}_1,{\bf r}_3)  h({\bf r}_3,{\bf r}_2) d{\bf
r}_3.
\label{p1}
\end{equation}  where $\rho({\bf r})$ is the inhomogeneous density, and, in
terms of functional derivatives, $h$ and $c^{(2)}$  are defined (fixed
$T$,$\mu$)  by
\begin{equation}
\beta^{-1}{\delta \rho({\bf r}_1) \over \delta V_{\text{ext}}({\bf r}_2)}=   -
\rho({\bf r}_1)\rho({\bf r}_2) h({\bf r}_1,{\bf r}_2) -\rho({\bf r}_{1})
\delta({\bf r}_{1}-{\bf r}_{2}) \,
\label{p2}
\end{equation} and
\begin{equation}
\beta{\delta V_{\text{ext}}({\bf r}_1)] \over \delta \rho({\bf r}_2)}=
 c^{(2)}({\bf r}_1,{\bf r}_2)  -{\delta({\bf r}_{1}-{\bf r}_{2}) \over
\rho({\bf
r}_{1})}\, ,
\label{p3}
\end{equation} with  $\beta=1/k_{\text{B}}T$. From (\ref{p1})-(\ref{p3}) one
easily obtains
\begin{equation}
\int  {\delta \rho({\bf r}_1) \over \delta V_{\text{ext}}({\bf r}_3)} {\delta
V_{\text{ext}}({\bf r}_3) \over \delta \rho({\bf r}_2)} d{\bf r}_3 =
\delta({\bf r}_{1}-{\bf r}_{2}) \, ,
\label{p4}
\end{equation} which expresses that, via the OZ equation, $h$ and
$c^{(2)}$ are (essentially) functional inverses. We note that this
functional-inversion statement is only possible if there is a one-to-one
relation between $\rho({\bf r})$ and
$V_{\text{ext}}({\bf r})$ (fixed $T$,$\mu$), so that the OZ equation can be
viewed as a consequence of this relation. This viewpoint is usual in DFT  of
classical fluids in the grand canonical ensemble (GCE) where the  one-to-one
relation between
$\rho({\bf r})$ and  $V_{\text{ext}}({\bf r})$ is a key theorem (see, e.g.,
\cite{Evans79}) and the  OZ equation follows as a corollary \cite{Evans92}.

We now focus our attention on the canonical ensemble (CE) where the number
of particles $N$ is fixed. This means that the CE inhomogeneous density
$\rho_N$ is normalized to $N$, i.e.,
\begin{equation}
\int
\rho_N({\bf r}) d{\bf r}  =N \, .
\label{p5}
\end{equation} Furthermore, from the usual definition of the two particle
density $\rho^{(2)}_N$ (see, e.g.,
\cite{Hansen}) it is direct to obtain
$\int \rho^{(2)}_N({\bf r}_1,{\bf r}_2)d{\bf r}_2= (N-1)\rho_N({\bf r}_1)$, and
thus, taking into account that $\rho^{(2)}_N({\bf r}_1,{\bf r}_2)  =\rho_N({\bf
r}_1)\rho_N({\bf r}_2) \Bigl( h_N({\bf r}_1,{\bf r}_2) +1\Bigr)$, the following
constraint for $h_N$ arises in the CE:
\begin{equation}
\int \rho_N({\bf r}_1) h_N({\bf r}_1,{\bf r}_2) d{\bf r}_1=-1 \, ,
\label{p6}
\end{equation} which, as pointed out by Ashcroft \cite{Ashcroft95}, in the
context of electronic systems leads to the notion of {\it exchange-correlation
hole}.

We remark that Eq.~(\ref{p6})  is a result of considering a  system in the CE
where the number of particles cannot fluctuate. An important consequence of
this equation is that the OZ equation (\ref{p1}) is no longer valid in the CE,
since it is inconsistent with (\ref{p6}). Furthermore, since the OZ equation is
associated to the one-to-one relation between $\rho({\bf r})$ and
$V_{\text{ext}}({\bf r})$ it would seem that there is not a CE counterpart of
this one-to-one relation, which would be of serious concern in DFT of the CE.
The goal of this Letter is to clarify this point and to introduce a CE-OZ
equation  similar to (\ref{p1}).

It is clear that, fixed $N$ and $T$, the CE density $\rho_N$ is determined by
the external potential $V_{\text{ext}}$, but this potential is not (strictly)
unique: any potential that differs from  $V_{\text{ext}}$ by an additive
constant leads to the same
$\rho_N$ and thus the inhomogeneous density is not uniquely determined by
the external potential.  This does not contradict the first Hohenberg-Kohn
theorem for DFT\cite{Kohn95} because it allows for this additive constant.
However,  this (strict) nonuniqueness means that the (Legendre) transform
from the functional variable
$V_{\text{ext}}$ to the functional variable $\rho_N$ is not invertible and
thus
${\delta \rho_N({\bf r}_1) / \delta V_{\text{ext}}({\bf r}_2)}$ is a singular
matrix, which in turn implies that Eq. (\ref{p4}) cannot hold for the pair
$V_{\text{ext}}$, $\rho_N$. We want to stress that this is due to the
fixed-$N$ constraint which implies that $0=\delta N/\delta
V_{\text{ext}}({\bf r}_2)=$ $\int {\delta  \rho_N({\bf r}_1) /
\delta V_{\text{ext}}({\bf r}_2)} d{\bf r}_1$ which shows that the matrix is
singular
\cite{nota}.  Our proposal for solving this difficulty is to `bypass' the
constraint by means of an extended variable space where instead of
$V_{\text{ext}}$ alone  we consider the set
$\{ V_{\text{ext}}, N \}$ and,  instead of $\rho_N$, the set $\{ \rho_N,
\lambda_N\}$ where
$\lambda_N$ will be defined later and identified with the Lagrange multiplier
related to the fixed-$N$ constraint. The role played by the different
variables
is  clarified by rebuilding the main DFT  results in this  extended space. Our
derivation is based on  Legendre transform approach of Ref.~\cite{Argaman00}.

The CE Helmholtz free energy $F$ is a functional of the external potential
$V_{\text{ext}}$ and a function of  the number of particles
$N$ and the temperature $T$, i.e., $F=F(T,N,[V_{\text{ext}}])$. The density
$\rho_N$ and the `chemical potential' $\lambda_N$ (different from $\mu$!)
are functions of $T$ and $N$  and functionals of $V_{\text{ext}}$, and are
obtained from the following derivatives:
\begin{equation}
\rho_N({\bf r},[V_{\text{ext}}])=
\Biggl({\delta F \over \delta V_{\text{ext}}({\bf r}) }\Biggr)_{N} \, ,
\label{p7}
\end{equation}
\begin{equation}
\lambda_N [V_{\text{ext}}]=
\Biggl({\partial F \over \partial N }\Biggr)_{V_{\text{ext}}}
\label{p8}
\end{equation} where the explicit dependence on $T$ has been been omitted.
We next consider the Legendre transform of $F$ to the new variables $\rho_N$
and $\lambda_N$,
\begin{equation} {\hat F} (\lambda_N,[\rho_N])=
 F (\lambda_N,[\rho_N]) -\int \rho_N({\bf r}) V_{\text{ext}}({\bf
r},\lambda_N,[\rho_N]) d{\bf r} -\lambda_N \int \rho_N({\bf r}) d{\bf r} \, ,
\label{p9}
\end{equation} where we have used $N=$ $\int \rho_N({\bf r}) d{\bf r}$
[Eq.~(\ref{p5})] and we define
$F (\lambda_N,[\rho_N])\equiv$ $F( \int \rho_N({\bf r}) d{\bf
r},[V_{\text{ext}}(\lambda_N,[\rho_N])])$. In order to introduce a variational
principle we `revert' this transform by obtaining a functional that depends on
both the new variables
$\{\rho_N,\lambda_N\}$ and the old ones
$\{ V_{\text{ext}},N\}$:
\begin{equation} F_{V_{\text{ext}}, N} (\lambda_N,[\rho_N])=
 {\hat F} (\lambda_N,[\rho_N]) +\int \rho_N({\bf r}) V_{\text{ext}}({\bf r})
d{\bf r}+\lambda_N N\, .
\label{p10}
\end{equation} The minimization of this functional w.r.t.
$\{\rho_N,\lambda_N\}$ leads to
\begin{equation}
\Biggl( {\delta {\hat F} (\lambda_N,[\rho_N])
\over
\delta \rho_N({\bf r}) }
\Biggr)_{\lambda_N}+V_{\text{ext}}({\bf r})=0 \,
\label{p11}
\end{equation} and
\begin{equation}
\Biggl( {\partial {\hat F} (\lambda_N,[\rho_N])
\over
\partial \lambda_N }
\Biggr)_{\rho_N}+N  =0\,.
\label{p12}
\end{equation} Substituting (\ref{p9}) into (\ref{p12}) and taking into account
Eq.~(\ref{p5}) one has that the functional
\begin{equation} {\cal F}\equiv {\hat F}+\lambda_N \int \rho_N({\bf r}) d{\bf
r} = F(\lambda_N,[\rho_N]) -\int \rho_N({\bf r}) V_{\text{ext}}({\bf
r},\lambda_N,[\rho_N]) d{\bf r}
\label{p12bis}
\end{equation}  does not depend on $\lambda_N$, that is ${\cal F}={\cal
F}[\rho_N]$ and thus, from (\ref{p11}) we obtain
\begin{equation}
\label{p13} {\delta  {\cal F}[\rho_N] \over \delta \rho({\bf
r})}\Biggr|_{\rho=\rho_{\text{gc}}}+ V_{\text{ext}}({\bf r}) =\lambda_N\,.
\end{equation} Comparing  this equation with the GCE Euler-Lagrange equation
we see that $\lambda_N$ plays the role of the chemical potential $\mu$ (see,
e.g., Ref.~\cite{Evans79}). If on the other hand we compare with the CE
Euler-Lagrange equation obtained in Ref.~\cite{white00} (see also
\cite{Ashcroft95}) by means of the Lagrange multiplier technique we can
identify $\lambda_N$  with the Lagrange multiplier associated to the
constraint (\ref{p5}) and ${\cal F}$ is the intrinsic free energy functional in
the CE. Therefore, we see  that using the extended variable space one obtains
the same  Euler-Lagrange equation than in the standard DFT procedure,  with a
intrinsic free energy functional ${\cal F}$ that only depends on the density.

As an example we apply the above expressions to the classical ideal gas. For
this system one has
$-\beta F(T,N,[V_{\text{ext}}])=N \log\Bigl( \Lambda^{-3}
\int \exp[-\beta V_{\text{ext}}({\bf r})]  d{\bf r} \Bigr)-\log N!$ where
$\Lambda$ is the thermal wavelength. Using  (\ref{p7})  we obtain the familiar
barometric result
$\rho_N({\bf r},[V_{\text{ext}}])=N \exp[-\beta V_{\text{ext}}({\bf r})] / \int
\exp[-\beta V_{\text{ext}}({\bf r})]  d{\bf r}$ and from (\ref{p8}) we would
trivially obtain the corresponding result for $-\beta\lambda_N
[V_{\text{ext}}]$. These results can be inverted to obtain
$N[\rho_N]=\int \rho_N({\bf r}) d{\bf r}$ and an explicit expression (not
shown) for
$V_{\text{ext}}$ in terms of $\rho_N$ and $\lambda_N$. Substituting these
expressions in  (\ref{p12bis}) we obtain the following result for the intrinsic
free energy functional of the ideal gas in the CE:
\begin{equation}
\beta{\cal F}_{\text{id}}[\rho_N]= -\int \rho_N({\bf r}) \Bigl( \log
\bigl(\Lambda^{-3}
\rho_N({\bf r})\bigr) -1\Bigr) d{\bf r} + \phi(\int \rho_N({\bf r})  d{\bf
r}) \,.
\label{p14}
\end{equation}  with $\phi(x)=\log x! -x\log x+x$. We note that this
free-energy only differs from the GCE result by the term  $\phi(\int
\rho_N({\bf r})  d{\bf r})$, i.e., by $\phi(N)$. For large
$N$,  $\phi(N)\approx (\log N)/2$, and, since the mean square fluctuation
$\Delta^2(N)$ of the ideal gas in the GCE is equal to $N$ we find that
(\ref{p14}) agrees with the saddle point approximation of
Ref.~\cite{white00}  in which the leading correction to the CE free energy
functional is given by $(\log [\Delta^2(N)])/2$. Finally, we recall that using
(\ref{p14}) in Eqs. (\ref{p11})-(\ref{p12bis}) we rederive the result for the
equilibrium density
$\rho_N$, as we would have done using  (\ref{p14}) and (\ref{p13}) with the
constraint (\ref{p5}).

An important property of using the extended variable space is that, as we have
seen in the example, the set
$\{\rho_N,\lambda_N\}$ determines completely the set  $\{
V_{\text{ext}},N\}$, and, since the opposite is also (trivially) true,
there is a
one-to-one relation between both sets. This means that the Lagrange
transform is invertible and the (Hessian) matrix $\partial (\rho_N,\lambda_N)
/\partial (V_{\text{ext}},N)$ is not singular. Multiplying this matrix by its
inverse
$\partial (V_{\text{ext}},N)/\partial (\rho_N,\lambda_N)$ and equating to the
identity matrix in the  extended space, we obtain a set of equations that, in
the canonical ensemble, correspond to the OZ equation. After some algebra we
obtain the exact relations
\begin{equation} h_N({\bf r}_1,{\bf r}_2) = c^{(2)}_N({\bf r}_1,{\bf r}_2) +
\int \rho_N({\bf r}_3) c^{(2)}_N({\bf r}_1,{\bf r}_3)  h_N({\bf r}_3,{\bf r}_2)
d{\bf r}_3   -{1\over \rho_N({\bf r}_2)}{\partial \rho_N({\bf r}_2)\over
\partial N}
\label{p15}
\end{equation}  and
\begin{equation} {\partial \rho_N({\bf r}_2)\over \partial N}= {\rho_N({\bf
r}_2)\over N}
\Bigl(1+\int\int
\rho_N({\bf r}_1){\partial \rho_N({\bf r}_3)\over \partial N}
\Bigl( c^{(2)}_N({\bf r}_2,{\bf r}_3)-c^{(2)}_N({\bf r}_1,{\bf r}_3)
\Bigr) d{\bf r}_1d{\bf r}_3\Bigr),
\label{p16}
\end{equation}  where in analogy with the GCE we have introduced the CE
direct correlation function $c^{(2)}_N$ as
\begin{equation} c^{(2)}_N({\bf r}_1,{\bf r}_2)= -\beta{\delta^2 ({\cal
F}[\rho_N]-{\cal F}_{\text{id}}[\rho_N])\over
\delta \rho_N({\bf r}_1)\delta \rho_N({\bf r}_2)}
\label{p17}
\end{equation} with  ${\cal F}_{\text{id}}$ given by (\ref{p14}). Eqs.
(\ref{p15})-(\ref{p17}) are the main results of the present work. These
equations show that it is possible to derive the CE counterpart of the direct
correlation function and the OZ equation. The new OZ equation is consistent
with the constraints (\ref{p5}) and (\ref{p6}) as one can readily show by
noting that, independently of the functional form of $c^{(2)}_N$, these
constraints lead to identities when used in (\ref{p15}) and (\ref{p16}) (after
an  appropriate integration).

Given an explicit form for the intrinsic free-energy functional $\cal F$, one
can obtain $c^{(2)}_N$ via functional differentiation and then ${\partial
\rho_N({\bf r}_2)/
\partial N}$ is obtained from (\ref{p16}). Finally, using (\ref{p15}) one
obtains the total correlation function $h_N$. In general, solving these
equations is not an easy task and one must resort to self-consistent
procedures. There are some cases in which one can obtain the required
information by means of simple methods. The simplest situation arises for
the ideal gas. In this case $c^{(2)}_N=0$, Eq. (\ref{p16}) implies
 ${\partial \rho_N({\bf r})/ \partial N}=$ ${\rho_N({\bf r})/ N}$ and, from
(\ref{p15}) we obtain the well-known result $h_N=-1/N$ for the total
correlation function of the ideal gas.

Another simple situation arises in the uniform limit. Considering that in this
limit
 $\rho_N({\bf r})
\to$
$\rho_N\equiv$
$N/V$ ($V$ is the volume of the system),  one has ${\partial \rho_N/ \partial
N}=$ ${\rho_N/ N}$ and thus Eq.~(\ref{p16}) becomes
\begin{equation} h_N(r_{12}) = c^{(2)}_N(r_{12}) +
\rho_N \int  c^{(2)}_N(r_{13})  h_N(r_{32}) d{\bf r}_3   -{1\over  N}
\label{p18}
\end{equation}  where $|{\bf r}_i-{\bf r}_j|\equiv r_{ij}$ and we have taken
into account that the uniform fluid is both translationally and rotationally
invariant.  This equation should be compared with the GCE-OZ equation
(\ref{p1}), that,  in the uniform limit,
$\rho({\bf r}) \to$ $\rho=$ $\rho_N$, reduces  to
\begin{equation} h(r_{12}) = c^{(2)}(r_{12}) +
\rho \int  c^{(2)}(r_{13})  h(r_{32}) d{\bf r}_3\,.
\label{p19}
\end{equation}  Then, for uniform fluids, both equations only differ by the
$1/N$ term which vanishes in the thermodynamic limit.

It is well-known \cite{Lebowitz61,Salacuse} that the GCE radial distribution
function
$g(r)=h(r)+1$ can be related to $g_N(r)=h_N(r)+1$ by means of a series
expansion in powers of
$1/N$. Using this expansion one has for the total correlation function (with
$\rho_N=$ $\rho$)
\begin{equation} h_N(r) \approx h(r) - {S(0) \over N} -{S(0)\over
2N}{\partial^2\over \partial \rho^2}\Biggl(\rho^2 [h(r) - {S(0) \over N}]
\Biggr)
\,,
\label{p20}
\end{equation}  where $S(k) \equiv 1+\rho \tilde{h}(k)$ is the structure
factor, being $\tilde{h}(k)=\int h(r)
\exp (i{\bf k}\cdot {\bf r})d{\bf r}$ the Fourier transform of $h(r)$. In what
follows we shall indicate how one can rederive this result by means of the OZ
equations (\ref{p18}) and (\ref{p19}) and an approximate expression for
$c_N^{(2)}$ in terms of $c^{(2)}(r)$ (conversely, Eq.~(\ref{p20}) and
Eqs.~(\ref{p18})-(\ref{p19}) can be used to obtain an approximate expression
for
$c_N^{(2)}$).  Since the OZ equations (\ref{p18}) and (\ref{p19}) involve
convolutions it is convenient to work in Fourier space. Subtracting (\ref{p19})
from  (\ref{p18}), after some algebra, we obtain the exact relation
\begin{equation}
\Delta\tilde{h}(k) = {-8\pi^{3}N^{-1}S(k)\delta(k) +\Delta\tilde{c}(k)S^2(k)
\over  1-\rho S(k)\Delta\tilde{c}(k)}
\label{p21}
\end{equation}  where $\Delta\tilde{h}(k)\equiv$ $\tilde{h}_N(k)
-\tilde{h}(k)$ and
$\Delta\tilde{c}(k)\equiv$ $\tilde{c}^{(2)}_N(k) -\tilde{c}^{(2)}(k)$. From
this
equation [or, alternatively, from Eq.~(\ref{p6})] one easily obtains
$\Delta\tilde{h}(0)=-S(0)/\rho$, i.e., although $\Delta{h}(r)$ is of order
$1/N$, its integral over the total volume $V$ does not vanish with increasing
$N$. A simple analysis of Eq.~(\ref{p21}) can be done by considering that,
since the difference between the CE and the GCE free-energy functionals is of
order $\log N$  \cite{white00}, one has that $\Delta \tilde{c}^{(2)}(k)$ is of
order $1/N$, and thus, to first order in $\Delta \tilde{c}^{(2)}(k)$ we obtain
\begin{equation}
\Delta\tilde{h}(k) \approx -{8\pi^{3}S(k)\delta(k) \over N}+
\Biggl({S(k)\over \rho}-{8\pi^{3}S(k)\delta(k) \over N}\Biggr)\rho
S(k)\Delta\tilde{c}(k)\,.
\label{p22}
\end{equation}  We note that, to zeroth order, this equation reduces to
$\Delta\tilde{h}(k) \approx$
$-{8\pi^{3}S(k)\delta(k) / N}$ and thus $\Delta{h}(r) \approx$ $-S(0)/ N$,
i.e.,
the first contribution to Eq.~(\ref{p20}). In order to evaluate (\ref{p22}) we
need an expression (valid to first order) for $\Delta {c}^{(2)}$, this
expression
can be obtained from the above-mentioned saddle point approximation for the
CE free-energy functional \cite{white00} which leads to
\begin{equation}
\Delta {c}^{(2)}(r_{12}) \approx -{1 \over 2} {\delta^2\over\delta\rho({\bf
r}_1)\delta\rho({\bf r}_2)}
\Biggl(\log{\Delta^2(N)\over N}\Biggr)\,.
\label{p23}
\end{equation}  From this approximation, a rather lengthy calculation shows
that Eq.~(\ref{p22}) becomes the  series expansion result of Eq.~(\ref{p20}).
This demonstration is not only a proof of the usefulness of the CE-0Z equation
but also it gives further support to the saddle point approximation of
Ref.~\cite{white00}. We note that this approximation has been shown
\cite{whiteb00} to yield results for the inhomogeneous density that are
equivalent to first order to those of the series expansion of the CE
inhomogeneous density introduced by Gonz\'alez {\it et al} \cite{Gonzalez97}.

In summary, we have seen that the fixed-$N$ constraint of the CE is
inconsistent with the standard OZ relation in the GCE. By resorting to DFT
methods we have identified the source of this inconsistency with the lack of a
strict one-to-one relation between the density and the external potential
which leads to a singular Hessian matrix. This problem has been solved by
means of an extended variable-space in which we rederive the usual CE
Euler-Lagrange equation with the advantage that the new Hessian matrix is no
longer singular and allows to  introduce a CE-OZ relation. For both the ideal
gas and the uniform fluid this  CE-OZ  relation only differs from GCE-OZ by a
$1/N$ term. The presence of this term leads to the exact total correlation
function for the ideal gas. For the uniform fluid we recover the result of the
standard series expansion of the CE total correlation function.

\acknowledgements We  thank financial support by the Comisi\'on
Interministerial de Ciencia y Tecnolog\'{\i}a of Spain under Grant PB 98-0261.


\begin{references}
\bibitem{nota}An equivalent demonstration follows from noting
 that (\ref{p2}) is also valid in the canonical ensemble, which together with
Eq.~(\ref{p6}) leads to
$-\beta^{-1}\int {\delta  \rho_N({\bf r}_1) /
\delta V_{\text{ext}}({\bf r}_2)} d{\bf r}_1=0$.
\end{references}
\end{document}